\documentstyle[12pt]{article}

\begin{document}

\hoffset = -1.2 truecm \voffset = -2.2 truecm \baselineskip 24pt

\begin{center}
\vskip 2cm
{\large {\bf On the Spin-Orbital Structure of the ``Zero'' Magnetization\\[0pt]
Spin-Precessing Modes of the Superfluid {\bf $^3He-B$}}}\\[0pt]

\bigskip

G. Kharadze and N. Suramlishvili

\bigskip
     
{\it Andronikashvili Institute of Physics, Georgian Academy of Sciences,
6 Tamarashvili St., 380077, Tbilisi, Georgia}\\[0pt]
 
\bigskip

\end{center}

{\small  The spin-orbital configurations of the coherently precessing
spin modes characterized by a small value of the magnetization
$(M\ll M_{0}=\chi H_{0})$ is considered. Various regimes are
analysed depending on the transverse rf-field strength.}

     
\bigskip

In the superfluid phases of liquid $^3He$ the unusual spin-precessing
modes can be exited at the magnitude $M=|\vec M|$ of the magnetization
essentially different from its equilibrium value $M_0=\chi H_0$
(here $\chi$ is the
magnetic susceptibility and $H_0$ denotes the strength of an applied
static magnetic field). Such a possibility is realized, in particular,
for $^3He-B$ at $M=M_0/2$ (or $M=2M_0$), as has been pointed out in Ref.[1]
and demonstrated experimentally [2-4] (for the details see, also, Ref.[5]).
These homogeneously precessing spin-modes are stabilized at the local minima
of time-averaged dipole-dipole potential $U_D$ and are characterized by the
specific (non-Leggett) orbital configurations of the order parameter.

For the superfluid $B$-phase
\begin{equation}
U_D=\frac{2}{15}\chi_B\left(\frac{\Omega_B}{g}\right)^2
\left(Tr\hat{R}-\frac{1}{2}\right)^2,
\end{equation}
where $\Omega_B$ is the longitudinal NMR frequency, $g$-the gyromagnetic
ratio for $^3He$ nuclei and the orthogonal matrix $\hat{R}$ describes 3D
relative rotations of the spin and orbital degrees of freedom. By
introducing the triples of Euler angles $(\alpha_S,~ \beta_S,~ \gamma_S)$
and $(\alpha_L,~ \beta_L,~ \gamma_L)$, which parametrize
rotations in the spin and orbital spaces, respectively, it can be shown that
$U_D=U_D(s_Z, l_Z, \alpha, \gamma)$, where $s_Z=\rm{cos}\beta_S
~(l_Z=\rm{cos}\beta_L)$
is the projection of the Cooper pair spin quantization axis (orbital momentum
quantization axis) along the direction of the applied static magnetic field
$\vec{H}_0,~ \alpha=\alpha_S-\alpha_L$ and $\gamma=\gamma_S-\gamma_L$.
Measuring the energy density in units of $\chi H_0^2$ and noticing that $U_D$
is a periodic function of $\alpha$ and $\gamma$, we have:

\begin{equation}
\frac{U_D}{\chi_B H_0^2}=\varepsilon f(s_Z,l_Z,\alpha,\gamma)=
\varepsilon \sum_{kl} f_{kl}(s_Z, l_Z)e^{i(k\alpha+l\gamma)}
\end{equation}
with $\varepsilon\propto(\Omega_B/\omega_0)^2$ (here $\omega_0=gH_0$).

In the case of a strong magnetic field ($\omega_0\gg\Omega_B$) the angular
variables $\alpha$ and $\gamma$ perform rapid rotations within a long time
scale $1/\Omega_D$ and the result of the time-averaging procedure of Eq.(2)
essentially depends on the possible presence of a slow combination
$\phi_{kl}=k\alpha+l\gamma$. Since in the strong-field case
($\varepsilon\ll 1$)
$\dot{\alpha}\simeq -\omega_0$ and $\dot{\gamma}\simeq gM/\chi$, the phase
$\phi_{kl}$ turns out to be a slow variable at the resonance condition
$M/M_0=k/l$. Addressing the explicit expressions of the Fourier
coefficients $f_{kl}$ for $^3He-B$ (see Ref.[6]) it is concluded that, along
with the conventional resonance at $M=M_0$ with $k=l=\pm 1,\pm 2$, the two
other (unconventional) resonances are realized at $M=M_0/2~(M=2M_0)$ with
$2k=l~(k=2l)$. This happens in the case of $l_Z\neq 1$ (non-Leggett orbital
configuration) where $f_{12}=f_{21}\neq 0$. In particular, at $M=M_0/2$ the
time-averaged dipole-dipole potential

\begin{equation}
\bar{f}=1+2s_Z^2 l_Z^2+(1-s_Z^2)(1-l_Z^2)+\frac{2}{3}
\sqrt{1-s_Z^2}\sqrt{1-l_Z^2}(1+s_Z)(1+l_Z)\rm{cos}\phi_{12}
\end{equation}
and the half-magnetization (HM) spin-precessing mode is trapped at the local
minima of Eq.(3) [1].

In Refs.[2,3], along with the above-mentioned HM spin-precessing mode, another
unusual spin-precessing state with $M\ll M_0$ has been observed experimentally.
This ``zero'' magnetization mode is not within the category
of the resonances with a slow phase. Instead, the ``zero'' magnetization
spin-precessing mode can be stabilized at the balance of the dissipative
energy losses and the transverse rf-field energy pumping, as has been discussed
in Ref.[7]. The experimentally realized ``zero'' magnetization spin-precessing
mode seems to be characterized by the Cooper pairs orbital configuration close
to $l_Z=0$, although in Ref.[7] the case of the Leggett configuration
($l_Z=1$) was adopted. The computer simulation was used to resolve this
controversy (see, e.g., [8]). Here we apply an analytical approach to the
same question.

In order to construct the stationary spin-precessing state with $M\ll M_0$
we address the equations describing the evolution of the spin density
$\vec {S}=\vec {M}/g$. In the strong magnetic field case the spin dynamics
is governed by the two pairs of canonically conjugate variables ($S_Z,\alpha$)
and ($S,\gamma$) subject to a set of the equations

\begin{equation}
\dot{\alpha}=-1+\varepsilon\frac{\partial f_D}{\partial S_Z}+
\frac{}{S_Z}{\sqrt{S^2-S_Z^2}}h_{\bot}\rm{cos}\theta-
\varepsilon\kappa\frac{S^2}{\sqrt{S^2-S_Z^2}}
\left(\frac{\partial f_D}{\partial\alpha}-
s_Z\frac{\partial f_D}{\partial\gamma}\right),\\
\end{equation}

\begin{equation}
\dot\gamma=S+\varepsilon\frac{\partial f_D}{\partial S}-
\frac{S}{\sqrt{S^2-S_Z^2}}h_{\bot}\rm{cos}{\theta}-
\varepsilon\kappa\frac{S^2}{\sqrt{S^2-S_Z^2}}
\left(\frac{\partial f_D}{\partial\gamma}-
s_Z\frac{\partial f_D}{\partial\alpha}\right),\\
\end{equation}

\begin{equation}
\dot{S_Z}=-\varepsilon\frac{\partial f_D}{\partial\alpha}-
\sqrt{S^2-S_Z^2}h_{\bot}\rm{sin}\theta+\varepsilon^2\kappa(S^2-S_Z^2)
\left(\frac{\partial f_D}{\partial S_Z}\right)^2,\\
\end{equation}

\begin{equation}
\dot{S}=-\varepsilon\frac{\partial f_D}{\partial\gamma}.
\end{equation}

This equations are put into a dimensionless form with the time measured is
the units of $1/\omega_0$ and the spin density - in the units of
$S_0=M_0/g$. The presence of the transverse rf field
$\vec{H}_{\bot}(t)=H_{\bot 0}(\hat{x}\rm{cos}\phi(t)+\hat{y}\rm{sin}\phi(t))$ is
taken into account $(h_{\bot}=H_{\bot 0}/H_0, \theta=\alpha-\phi)$.
The dissipation in the spin dynamics is characterized by a phenomenological
parameter $\kappa$ (as in Ref.[9]).

Now we focus on the case of the spin precession in the regime with
$S\ll 1$. In this situation $\dot\gamma\ll 1$ but if
$\sqrt{\varepsilon}\ll S\ll 1$ the angular variable will change faster
then $S_Z$ and $S$. Then, in order to describe the evolution of the
spin system in $\varepsilon$-approximation (the Van der Pol picture),
we have to use the time-averaged dipole-dipole potential $\bar{f}_D$
with respect to both angular variables $\alpha$ and $\gamma$ independently.
In this non-resonance case

\begin{equation}
\bar{f}=f_{00}=1+2s_{Z}^{2}l_{Z}^{2}+
(1-s_{Z}^{2})(1-l_{Z}^{2}),
\end{equation}
and the disipationless dynamics of $\alpha$ and $\gamma$ is governed by
the equations

\begin{equation}
\dot{\alpha}=-1+\varepsilon \frac{\partial f_{00}}{\partial S_Z}+
\frac{S_Z}{\sqrt{S^2-S_{Z}^2}} h_{\bot} \rm{cos}\theta,\\
\end{equation}
\begin{equation}
\dot{\gamma}=S+\varepsilon \frac{\partial f_{00}}{\partial S}-
\frac{S}{\sqrt{S^2-S_{Z}^2}} h_{\bot}\rm{cos}\theta,
\end{equation}
with $S_Z$ and $S$ being constants.

The stationary solutions of Eqs.(9) and (10) are found according to the
conditions

\begin{equation}
\frac{\partial{f}}{\partial S_Z}=\frac{\partial f}{\partial S} =0,
\end{equation}
where the free energy

\begin{equation}
f=\varepsilon f+\frac{1}{2}(S-\omega_{\gamma})^2+(\omega_{\alpha}-1)S_{Z}-
\sqrt{S^2-S_Z^2}h_{\bot} \rm{cos}\theta
\end{equation}
with $\omega_{\alpha}=-\dot{\alpha}|_{st}$ and
$\omega_{\gamma}=\dot{\gamma}|_{st}$.

In what follows we consider the case where the first (dipole-dipole) term
in Eq.(12) dominates over the rest part of the free energy. Than
it is concluded that the stationarity conditions (11) are reduced to the
equation

\begin{equation}
\frac{\partial f_{00}}{\partial\beta_s}=2\sqrt{1-s_Z^2}s_Z (1-3 l_Z^2)=0.
\end{equation}

In the situation where $l_Z$ is free to adjust to the minimal value of
$f_{00}$, Eq.(13) is to be supplemented by an analogous condition

\begin{equation}
\frac{\partial f_{00}}{\partial\beta_L}=2\sqrt{1-l_Z^2}l_Z (1-3 s_Z^2)=0.
\end{equation}

From Eqs.(13) and (14) it is readily concluded that the minimal value
of $f_{00}(=1)$ is realized at the following spin-orbital configurations:

\begin{eqnarray}
a)~~s_Z=\pm 1,~~~l_Z=0,\\
b)~~s_Z=0,~~~l_Z=\pm 1.
\end{eqnarray}
The case with $s_Z=l_Z=1$ corresponds to the maximal value of $f_{00}$.

The experimentally observed ``zero'' magnetization spin-precessing mode
is developed near the spin-orbital configuration $s_Z=1$ and $l_Z=0$ [2,3]
and we have to verify whether  it corresponds to the dynamical regime
with $\sqrt{\varepsilon}<S\ll 1$. For this purpose the set of equations
for $S$ and $S_Z$ with dissipative terms is to be addressed. Using the
results of Ref.[6] and puting $s_Z\simeq 1-\frac{1}{2}\beta_S^2$, for
the case $l_Z=0$ the following set of equations is obtained:

\begin{eqnarray}
\dot{S}=4\kappa\varepsilon^2\left(1-\frac{\beta_S^2}{2S}\right),\\
\dot{\beta_S}=-\frac{\kappa\varepsilon^2}{S\beta_S}
\left(\frac{77}{9}+\frac{\beta_S}{S}\right)+h_{\bot} \rm{sin}\theta,
\end{eqnarray}
with the stationary solution

\begin{equation}
S=a\left(\frac{\varepsilon^2}{h_{\bot}}\right)^{2/3}.
\end{equation}
where $a=\frac{1}{2}\left(\frac{190\kappa}{9sin\theta}\right)^{2/3}$.
It can be verified that this solution is locally stable. Since in Eq.(19)
the coefficient $a$ is of the order of unity, the initially imposed condition
on the value of $S~(\sqrt{\varepsilon}\ll S \ll 1)$ will be fulfilled for
$h_{\bot}$ being within the limits

\begin{equation}
\varepsilon^2\ll h_{\bot} \ll \varepsilon^{5/4}.
\end{equation}

For the case with $\varepsilon\simeq 10^{-3}$, wich corresponds to the
situation realized experimentally, the value of
$h_{\bot}=H_{\bot}/H_0\simeq 10^{-5}$ fits Eq.(20).
This consideration shows that the ``zero'' magnetization
spin-precessing mode is realized at the spin-orbital configuration
$s_Z\simeq 1,~l_Z\simeq 0$, in accordance with experimental observations.

Now we turn to the quastion of stability of the ``zero''
magnetization spin-precessing mode near the orbital state $l_{Z}=1$,
which realizes the maximum value of the time-averaged dipole-dipole
potenyial (8). Again addresing the set of Eqs. for $S$ and $s_Z$
(see Ref.[6]) this time for the case of a spin-orbital configuration
with $s_{Z}\simeq 1$ and $l_{Z}\simeq 1$, we obtain that for
$\sqrt{\varepsilon}\ll S \ll 1$

\begin{equation}
\dot{S}=\frac{160}{9}\kappa\varepsilon^{2},
\end{equation}
\begin{equation}
\dot{\beta_S}=-\frac{\kappa\varepsilon^2}{S\beta_S}
\left(\frac{77}{9}+\frac{\beta_S}{S}\right)+h_{\bot}\rm{sin}\theta.
\end{equation}

It can be easily seen that this set of Eqs. has no stationary solutions.

It is interesting to analyze another possible regime of ``zero''
magnetization spin-precessing state with $S\simeq\sqrt{\varepsilon}$.
In this case $\dot\gamma\simeq\sqrt{\varepsilon}$ and the rate of the
time evolution of angular variable $\gamma$ is still faster then the
temporal variations of $S$ and $S_Z$. On the other hand,
$\dot{s_Z}=(\dot{S_Z}-s_{Z}\dot{S})/S\simeq\sqrt{\varepsilon}$ and the
only fast variable upon which depends the dipole-dipole potential is
$\alpha$. In the considered situation

\begin{equation}
f_D=\sum_{k}f_{k}(s_Z,~l_Z,~\gamma)e^{ik\alpha},
\end{equation}
where the time-averaged value

\begin{equation}
f_0=1+2s_Z^2 l_Z^2+4s_Z l_Z \sqrt{1-s_Z^2}\sqrt{1-l_Z^2} \rm{cos}\gamma+\\
(1-s_Z^2)(1-l_Z^2)(1+\rm{cos} 2\gamma).
\end{equation}

After having minimized $f_0$ with respect to the slow variable $\gamma$
it can be shown that $f_0 (s_Z,~l_Z)$ is highly degenerate with respect to
the spin and orbital variables: the minimum of $f_0 (s_Z,~l_Z)$ is
realized within the circle $s_Z^2+l_Z^2\leq 1$. Considering again the
orbital state $l_Z=0$ with the stationary value $\gamma_{st}=\pi/2$
it can be shown that the time evolution of $s_Z$ is governed by an
equation

\begin{equation}
\dot{s}_Z=\frac{4\kappa\varepsilon^2}{S(1-s_Z^2)}
(s_Z^4-s_Z^3+s_Z^2+4)-h_{\bot}\sqrt{1-s_Z^2}\rm{sin}\theta.
\end{equation}

From Eq.(25) it follow that near $s_Z=\pm 1$ the stationary value of
$\beta_S$ is given by

\begin{equation}
\beta_S \simeq 
2\left(\frac{2\kappa\varepsilon^2}{3S}
\frac{1}{h_{\bot}\rm{sin}\theta}\right)^{1/3}.
\end{equation}

It is easy to show that the solution (26) for the case $s_Z\simeq -1$
is locally stable (in contrast to the case with $s_Z\simeq 1$). This
spin-orbital configuration $(s_Z\simeq -1, l_Z\simeq 0)$ is one of the
stable spin-precessing states at $S\simeq\sqrt{\varepsilon}$. The phase
diagram of the ``zero'' magnetization in the degeneracy domain
$s_Z^2+l_Z^2\leq 1$ will be presented in a separate publication.

We are indepted to E.Sonin for his valuable comments.

\end{document}